\documentclass[11pt]{article}

\usepackage[margin=1in]{geometry}
\usepackage[T1]{fontenc}
\usepackage{times}
\usepackage{amsmath,amssymb,amsthm,mathtools}
\usepackage[authordate,backend=b
iber]{biblatex-chicago}
\usepackage{microtype}
\usepackage{hyperref}
\usepackage[nameinlink,noabbrev]{cleveref}
\usepackage{titlesec}

\hypersetup{colorlinks=true,linkcolor=blue,citecolor=blue,urlcolor=blue}
\titleformat*{\section}{\large\bfseries}
\titleformat*{\subsection}{\bfseries}
\titleformat*{\subsubsection}{\bfseries}

\newtheorem{theorem}{Theorem}
\newtheorem{proposition}[theorem]{Proposition}

\newtheorem{corollary}[theorem]{Corollary}
\theoremstyle{definition}

\theoremstyle{remark}
\newtheorem{remark}[theorem]{Remark}

\newcommand{\E}{\mathbb E}
\newcommand{\R}{\mathbb R}
\newcommand{\KL}{D_{\mathrm{KL}}}
\newcommand{\I}{I}
\newcommand{\bd}[1]{\boldsymbol{#1}}
\newcommand{\calQ}{{\cal Q}}
\newcommand{\calV}{{\cal V}}

\title{Rational Inattention to States and Choice Characteristics}
\author{Christopher W. Engh\footnote{Yale University. Email: \texttt{christopher.wu@yale.edu}}}
\date{}

\begin{document}
\maketitle
\vspace{-0.35in}

\begin{abstract}
This paper characterizes rationally inattentive behaviour under Shannon information costs when preferences depend on uncertain choice characteristics in addition to a payoff-relevant state. The optimal conditional choice probability exists, is unique, and assigns positive probability to every ambient characteristic, allowing dominated realized outcomes without treating them as deliberate choices. Conditional choice is pulled toward ubiquitous characteristics when options are costly to identify and toward unconditional behavior when states are costly to learn. A designer can improve welfare without estimating utility by shifting exposure toward observed demand. Compared to classic rational inattention, our model introduces only one new parameter, which is non-parametrically over-identified under product entry.

\end{abstract}

\section{Introduction}

Making a good choice often requires learning two different things: what the available options are and which option fits one's circumstances. A consumer choosing health insurance, for example, may be uncertain both about what each plan covers and about which health risks will matter. These two uncertainties call for different responses. Clearer plan descriptions can help a consumer who does not understand the contracts, but they do not resolve uncertainty about future health. To diagnose information frictions or design an effective intervention, an economist would therefore like to know which type of uncertainty is driving behavior.

Standard rational inattention does not generally make this distinction. It measures the information contained in the agent's eventual signal or action recommendation about the entire environment. But a short recommendation can depend on several uncertain objects at once. An advisor might tell the consumer which plan is appropriate without separately revealing either the consumer's future health or the plan's coverage. The recommendation can then be just as informative, in the Shannon sense, as a recommendation made to someone who already knows their health. The standard cost measures how much the final message reveals, not how much work was required to learn the different inputs used to construct it.

This paper develops a model that keeps these inputs separate. This requires departing from the usual rational-inattention setup, in which the alternatives and their characteristics are known and only the payoff-relevant state is uncertain. Simply adding an unknown menu to that state and charging for one final signal would recreate the original problem: it would not reveal whether the agent paid to identify the options, the state, or their relation.

I instead use a design-based ``Monty Hall'' environment. A fixed set of characteristics is assigned to otherwise arbitrary choice labels, but the assignment is initially unknown. As with prizes behind doors, the objects are not redrawn; uncertainty concerns which object is behind which label. Randomizing over compatible assignments makes arbitrary labels symmetric and isolates the information required to locate a characteristic from the information required to learn when it is valuable. The unusual setup is therefore a device for separating two economically different learning tasks, not an assumption that markets literally reshuffle products.

This formulation also has an exact reduced form. Under label symmetry and the requirement that the policy not collect payoff-irrelevant information, the agent can be represented as choosing a joint distribution over the characteristic ultimately selected and the payoff-relevant state. The reference distribution $\phi$ describes the characteristics the agent would encounter before payoff-directed information acquisition. In a literal menu, it is the fraction of labels carrying each characteristic. More generally, it can represent attention-weighted exposure, incorporating availability, placement, or salience. Thus the design both microfounds the characteristic-learning cost and produces a tractable object that a market designer may be able to influence.

The model produces a simple choice rule
\begin{align}\label{eq:ccp-intro}
p(x\mid\theta)
=
\frac{\phi(x)^\eta p(x)^{1-\eta}e^{u(x,\theta)/\lambda}}
{\sum_{y\in X}\phi(y)^\eta p(y)^{1-\eta}e^{u(y,\theta)/\lambda}}
\end{align}
where $p(x\mid\theta)$ is the probability of selecting characteristic $x$ in state $\theta$, $p(x)$ is its unconditional probability, and $u(x,\theta)$ is utility. In the filtering-and-sorting interpretation developed below, $\lambda$ is the unit price of sorting characteristics across states and $\eta\lambda$ is the unit price of filtering characteristics from the ambient distribution.

The rule combines two economic intuitions. First, if characteristics are costly to learn, a characteristic should be chosen more often when it becomes more ubiquitous in the ambient environment. The reason is simple: a common characteristic is easier to encounter and requires less filtering to select. Accordingly, holding utility and the marginal term fixed, an increase in $\phi(x)$ raises the conditional odds of selecting $x$. Second, standard rational inattention implies that state-conditional choice probabilities should be pulled toward their unconditional marginals when the state is costly to learn. Making $p(x\mid\theta)$ differ from $p(x)$ requires the agent to acquire information about $\theta$; when that information is expensive, behavior changes less across states. The model combines these forces by regularizing choice toward ambient ubiquity through $\phi$ and toward state-independent behavior through $p$. Their relative importance is governed by $\eta$. Thus observed behavior can reveal whether the agent is primarily struggling to identify the options or to identify the circumstances in which they are valuable.

The paper's first set of results establishes that the model is well behaved. Whenever learning characteristics has positive cost, an optimal information policy exists and is unique. Moreover, the agent never completely rules out a characteristic that has positive ambient exposure. Even a strictly dominated characteristic is selected with positive probability. This is not a claim that the agent knowingly chooses a dominated option. The primitive choice is a label, and the characteristic behind that label remains imperfectly known. Dominated outcomes arise because completely eliminating them would require costly information.

The second result concerns market or platform design. Ambient exposure can often be changed directly: a platform chooses which products enter a candidate set and how often they appear, a retailer controls shelf space and displays, and a comparison site controls the order and frequency with which plans are presented. In such settings the designer should increase the exposure of a characteristic if it is chosen more often than it is encountered, and decrease exposure if the reverse is true. This shifts $\phi$ toward the observed marginal distribution $p$. The rule requires only exposure and choice shares, which may be observed or experimentally varied. It avoids estimating utilities, information costs, and preference heterogeneity---objects that are often difficult to separate and sensitive to functional-form assumptions. The logic comes from the agent's own optimization. Observed choices already reflect the best tradeoff between utility and information costs. By making the revealed choice distribution easier to encounter, the designer can implement the same behavior with less information acquisition; allowing the agent to reoptimize can only add to the gain.

The third result provides a way to identify the relative information cost. Consider the entry of a new characteristic that leaves the relative ambient exposure of two incumbent characteristics unchanged. The change in their state-specific choice odds is tied to the change in their unconditional choice odds by a coefficient that identifies $\eta$. Repeating this comparison across states, incumbent pairs, or entry events generates over-identifying restrictions. This first step is useful because $\eta$ and the unobserved reference distribution $\phi$ are otherwise difficult to separate. After estimating $\eta$, the remaining information-cost, preference-heterogeneity, and exposure parameters can be estimated using standard discrete-choice methods.

The model builds on Shannon rational inattention \parencite{sims2003implications,mackowiak2023rational} and its discrete-choice formulation \parencite{matvejka2015rational,caplin2019rational}. The case in which only state information is costly recovers the familiar rational-inattention logit, while the case in which only characteristics are costly gives a reference-weighted logit. The paper is also related to work that enriches Shannon costs to capture the structure of the underlying information problem \parencite{fosgerau2020discrete,hebert2021neighborhood,caplin2022rationally,pomatto2023cost}. The contribution here is to isolate two economically distinct sources of uncertainty while retaining a tractable choice rule, a direct welfare prescription, and empirically testable restrictions.

The remainder of the paper derives the reduced problem from the finite menu model, establishes the choice and welfare results, and then discusses identification and estimation.

\section{Model}\label{sec:model}

\subsection{A finite Monty Hall environment}

Standard rational inattention begins with known actions and an uncertain payoff state. That starting point cannot separately price discovering what an action contains: the action's characteristics are already part of its name. The design below instead holds the menu fixed and randomizes only the mapping from characteristics to labels. This makes ambient frequencies independent of tastes, preserves symmetry across meaningless labels, and permits an exact reduction from label-level information policies to choice probabilities over characteristics.

Let the action labels be $A=\{1,\ldots,J\}$, the characteristic set $X$ be finite, and the state set $\Theta$ be finite. The menu contains $n_x$ copies of characteristic $x$, where
\begin{align*}
n_x\geq 0
\qquad
\sum_{x\in X}n_x=J
\qquad
\phi(x)=\frac{n_x}{J}
\end{align*}
Characteristics with $n_x=0$ can be omitted, so we take $\phi(x)>0$ without loss. Let $\calV(n)\subset X^J$ be the set of distinct permutations of this multiset and let
\begin{align*}
N=|\calV(n)|=\frac{J!}{\prod_{x\in X}n_x!}
\end{align*}
The random vector $\bd\xi=(\xi_1,\ldots,\xi_J)$ is uniform on $\calV(n)$. The state has full-support prior $\mu\in\Delta\Theta$, and
\begin{align}\label{eq:primitive-independence}
\bd\xi\mathrel{\perp\!\!\!\perp}\theta
\end{align}

The random assignment may be literal, as in an experiment, or it may represent the agent's symmetry before inspection. It does not require the agent to believe that Nature redraws products. The market is fixed; only the agent's information about the mapping is incomplete.

Let $\rho\in\Delta\calV(n)$ be the uniform design prior, so $\rho(\bd x)=1/N$. An information policy is a joint probability measure
\begin{align*}
\mathbb P\in\Delta(A\times\calV(n)\times\Theta)
\end{align*}
satisfying Bayes plausibility
\begin{align}\label{eq:primitive-bp}
\mathbb P_{\bd\xi,\theta}=\rho\otimes\mu
\end{align}
The recommendation is a label $a$, the resulting characteristic is $\xi=\xi_a$, and utility is $u(\xi_a,\theta)$. Thus utility depends on the characteristic behind the recommended label, not on the arbitrary label itself.

\subsection{Why mutual information with the action is not enough}

The standard Shannon cost would be $\lambda\I_{\mathbb P}(a;(\bd\xi,\theta))$. It asks how surprising the action recommendation is after observing the full environment. It does not ask whether constructing that recommendation required learning the design, the state, or only their relation. For example, $a$ can be independent of $\bd\xi$ and independent of $\theta$ while remaining informative about their joint realization. Conditioning on $a$ can then make the initially independent variables $\bd\xi$ and $\theta$ dependent. That induced dependence is exactly the information hidden by the one-message accounting.

We therefore use the primitive cost
\begin{align}\label{eq:primitive-cost}
\widehat C_{\eta,\lambda}(\mathbb P)
=
\eta\lambda\I_{\mathbb P}(a;(\bd\xi,\theta))
+
(1-\eta)\lambda\I_{\mathbb P}(\bd\xi;\theta\mid a)
\qquad
\eta\in[0,1]
\qquad
\lambda>0
\end{align}
The first term charges for resolving which label carries the useful characteristic. The second charges for the state--design relation that remains after a label has been recommended. When the state is already known, the second uncertainty disappears. When characteristics are already known, the chosen characteristic itself becomes the action and the second term becomes the standard state-information cost.

The probability accounting clarifies why this extension is parsimonious. The two information terms are
\begin{align}
\I_{\mathbb P}(a;(\bd\xi,\theta))
&=
\E_{\mathbb P}\left[
\log\frac{\mathbb P(a,\bd\xi,\theta)}
{\mathbb P(a)\mathbb P(\bd\xi,\theta)}
\right]
\label{eq:mi-first}
\\
\I_{\mathbb P}(\bd\xi;\theta\mid a)
&=
\E_{\mathbb P}\left[
\log\frac{\mathbb P(a,\bd\xi,\theta)\mathbb P(a)}
{\mathbb P(a,\bd\xi)\mathbb P(a,\theta)}
\right]
\label{eq:mi-second}
\end{align}
Bayes plausibility fixes $\mathbb P(\bd\xi)$, $\mathbb P(\theta)$, and $\mathbb P(\bd\xi,\theta)$. Arbitrary labels imply the symmetric choice $\mathbb P(a)=1/J$. Since labels themselves have no state-dependent payoff, a policy should not use them to signal the state beyond what is conveyed by the chosen characteristic; symmetry then gives $\mathbb P(a,\theta)=\mu(\theta)/J$. Under these restrictions, the only endogenous log probabilities in \cref{eq:mi-first,eq:mi-second} are
\begin{align*}
\log\mathbb P(a,\bd\xi,\theta)
\qquad
\log\mathbb P(a,\bd\xi)
\end{align*}
Indeed, up to a constant independent of the policy,
\begin{align}\label{eq:two-logs}
\frac{\widehat C_{\eta,\lambda}(\mathbb P)}{\lambda}
=
\E_{\mathbb P}\left[
\log\mathbb P(a,\bd\xi,\theta)
-(1-\eta)\log\mathbb P(a,\bd\xi)
\right]
+\text{constant}
\end{align}

These restrictions can be stated as three principles for an admissible optimum.

\begin{enumerate}
\item \textit{Label neutrality.} Relabeling the menu and the recommendation together does not change the policy
\item \textit{No state signaling through labels.} $\mathbb P(a,\theta)=\mathbb P(a)\mu(\theta)$
\item \textit{No irrelevant design information.} Conditional on $(a,\xi_a,\theta)$, the unchosen characteristics are uniform over all compatible assignments
\end{enumerate}

The first principle follows by symmetrizing any label-invariant problem. The latter two say that an optimal policy does not pay to encode information that neither changes the chosen characteristic nor affects utility. They are the information-acquisition analogue of discarding payoff-irrelevant signals.

\subsection{Reduction to chosen characteristics}

Let $P\in\Delta(X\times\Theta)$ denote a candidate distribution of the chosen characteristic and state, with $P_\theta=\mu$ and $P_\xi\ll\phi$. For any $\bd x\in\calV(n)$, write $x_a$ for its $a$th component.

\begin{theorem}[Equivalence]\label{th:equivalence}
There is a bijection between feasible reduced distributions $P$ over $(\xi,\theta)$ and primitive probability measures satisfying the three restrictions above. The primitive measure induced by $P$ is
\begin{align}\label{eq:canonical-policy}
\mathbb P_P(a,\bd x,\theta)
=
\frac{P(x_a,\theta)}{Nn_{x_a}}
\end{align}
It satisfies $\mathbb P_P(a)=1/J$ and $\mathbb P_P(a,\theta)=\mu(\theta)/J$. Expected utility is preserved and
\begin{align}\label{eq:cost-equivalence}
\widehat C_{\eta,\lambda}(\mathbb P_P)
=
\lambda\left[
\eta\KL(P_\xi\Vert\phi)
+
\KL(P_{\xi,\theta}\Vert P_\xi\otimes\mu)
\right]
\end{align}
Consequently, the primitive and reduced problems have the same unique distribution of observables whenever the reduced optimum is unique.
\end{theorem}

\begin{proof}
For any label $a$ and characteristic $x$, the number of assignments with $x_a=x$ is
\begin{align*}
|\{\bd x\in\calV(n):x_a=x\}|
=
\frac{(J-1)!}{(n_x-1)!\prod_{y\neq x}n_y!}
=
\frac{Nn_x}{J}
\end{align*}
Label neutrality assigns $P(x,\theta)/J$ to each label, while no irrelevant information spreads that mass uniformly over the $Nn_x/J$ compatible assignments. This yields \cref{eq:canonical-policy} and proves uniqueness. Summing \cref{eq:canonical-policy} over labels and assignments verifies that it induces $P$, while summing over labels verifies Bayes plausibility
\begin{align*}
\sum_{a=1}^J\mathbb P_P(a,\bd x,\theta)
=
\frac{1}{N}\sum_{x\in X}P(x,\theta)
=
\frac{\mu(\theta)}{N}
\end{align*}
The same counting gives
\begin{align*}
\mathbb P_P(a)
&=
\frac1J
\\
\mathbb P_P(a,\theta)
&=
\frac{\mu(\theta)}J
\\
\mathbb P_P(a,\bd x)
&=
\frac{P_\xi(x_a)}{Nn_{x_a}}
\end{align*}

The first likelihood ratio in \cref{eq:mi-first} therefore becomes
\begin{align*}
\frac{\mathbb P_P(a,\bd x,\theta)}
{\mathbb P_P(a)\rho(\bd x)\mu(\theta)}
=
\frac{P(x_a,\theta)}{\phi(x_a)\mu(\theta)}
\end{align*}
Because $(x_a,\theta)$ has distribution $P$ under $\mathbb P_P$, it follows that
\begin{align*}
\I_{\mathbb P_P}(a;(\bd\xi,\theta))
&=
\KL(P\Vert\phi\otimes\mu)
\\
&=
\KL(P_\xi\Vert\phi)+\I_P(\xi;\theta)
\end{align*}
The second likelihood ratio becomes
\begin{align*}
\frac{\mathbb P_P(a,\bd x,\theta)\mathbb P_P(a)}
{\mathbb P_P(a,\bd x)\mathbb P_P(a,\theta)}
=
\frac{P(x_a,\theta)}{P_\xi(x_a)\mu(\theta)}
\end{align*}
and hence
\begin{align*}
\I_{\mathbb P_P}(\bd\xi;\theta\mid a)
=
\I_P(\xi;\theta)
\end{align*}
Substitution into \cref{eq:primitive-cost} proves \cref{eq:cost-equivalence}. Expected utility is preserved because the induced law of $(x_a,\theta)$ is $P$.
\end{proof}

The theorem shows that the reduced problem is not a shortcut obtained by discarding a latent discrete-choice model. It is an exact representation of the observable implications of the primitive model under economically interpretable restrictions on useless information.

\subsection{The reduced problem}

The agent therefore solves
\begin{align}\label{eq:reduced-problem}
\max_{P\in\calQ}
\left\{
\E_P[u(\xi,\theta)]
-
\lambda\left[
\eta\KL(P_\xi\Vert\phi)
+
\I_P(\xi;\theta)
\right]
\right\}
\end{align}
where
\begin{align*}
\calQ
=
\{P\in\Delta(X\times\Theta):P_\theta=\mu\}
\end{align*}
and $P_\xi\ll\phi$ is automatic after restricting $X$ to the support of $\phi$. By the chain rule for relative entropy, the cost has the equivalent form
\begin{align}\label{eq:two-ways-cost}
C_{\eta,\lambda}(P)
&=
\eta\lambda\E_{\theta\sim\mu}
\left[
\KL(P_{\xi\mid\theta}\Vert\phi)
\right]
+
(1-\eta)\lambda\I_P(\xi;\theta)
\\
&=
\eta\lambda\KL(P_\xi\Vert\phi)
+
\lambda\I_P(\xi;\theta)
\end{align}
The first line retains the primitive Monty Hall accounting. The second line separates the reduced cost into an ambient-filtering term and a state-sorting term. This second decomposition is the basis for the independent axiomatic microfoundation below.

\subsection{Filtering and sorting as a separate microfoundation}\label{sec:filtering}

The finite-menu construction is one microfoundation for the reduced problem. A second construction starts directly from the two operations represented in \cref{eq:two-ways-cost}. Imagine a unit mass of particles, each carrying a characteristic $x\in X$. The ambient stock has composition $\phi$. The agent first filters this stock until the composition of the surviving particles is the desired marginal $p=P_\xi$. The agent then places those particles into boxes indexed by $\theta$. Box $\theta$ has capacity $\mu(\theta)$, and its desired composition is
\begin{align*}
q_\theta(x)
=
P(\xi=x\mid\theta)
\end{align*}
The capacity and material-balance restrictions are
\begin{align}\label{eq:sorting-balance}
P(x,\theta)
&=
\mu(\theta)q_\theta(x)
\\
p(x)
&=
\sum_{\theta\in\Theta}\mu(\theta)q_\theta(x)
\end{align}
Thus filtering changes the aggregate composition from $\phi$ to $p$, while sorting preserves the aggregate composition $p$ and changes only how characteristics are allocated across boxes.

\subsubsection{Axioms for filtering}

Let $b,q\in\Delta X$, with $q\ll b$, denote a baseline and a target composition. Write $\mathsf F(q\mid b)$ for the filtering cost per unit of delivered output. The following axioms describe an idealized filtering technology.

\begin{enumerate}
\item \textit{No creation.} If $q(x)>0$ while $b(x)=0$, then $\mathsf F(q\mid b)=+\infty$. A filter may discard or retain particles, but it cannot produce a type absent from the ambient stock at finite cost

\item \textit{Relative-frequency locality and output accounting.} There is a type-independent function $\ell_{\mathrm F}:\mathbb R_{++}\to\mathbb R$ such that
\begin{align}\label{eq:filter-locality}
\mathsf F(q\mid b)
=
\sum_{x:q(x)>0}q(x)
\ell_{\mathrm F}\left(\frac{q(x)}{b(x)}\right)
\end{align}
The relevant statistic for a delivered particle of type $x$ is its \emph{enrichment factor} $q(x)/b(x)$. Costs are accounted for per delivered particle and then averaged under the delivered composition $q$

\item \textit{Sequential consistency.} If a particle is enriched by a factor $r$ in one stage and by a factor $s$ in a second stage, the two-stage charge equals the direct charge
\begin{align}\label{eq:filter-composition}
\ell_{\mathrm F}(rs)
=
\ell_{\mathrm F}(r)
+
\ell_{\mathrm F}(s)
\end{align}
This is a path-independence requirement. Passing from $b$ to an intermediate composition $m$ and then from $m$ to $q$ should give the same charge to a delivered type as passing directly from $b$ to $q$, because
\begin{align*}
\frac{q(x)}{b(x)}
=
\frac{q(x)}{m(x)}
\frac{m(x)}{b(x)}
\end{align*}

\item \textit{Regularity.} The function $\ell_{\mathrm F}$ is continuous. Arbitrarily small changes in an enrichment factor cannot cause a discrete jump in cost

\item \textit{Normalization and aggregate nonnegativity.} Leaving the ambient composition unchanged costs zero, and no feasible target has negative total cost
\begin{align*}
\mathsf F(b\mid b)
&=
0
\\
\mathsf F(q\mid b)
&\geq
0
\end{align*}
If every nontrivial change is costly, the second inequality is strict whenever $q\neq b$
\end{enumerate}

\begin{proposition}[Axiomatic filtering cost]\label{prop:axiomatic-filtering}
The filtering axioms imply that there is a constant $\lambda_{\mathrm F}\geq0$ such that
\begin{align}\label{eq:axiomatic-filtering-cost}
\mathsf F(q\mid b)
=
\lambda_{\mathrm F}\KL(q\Vert b)
\end{align}
for every $q\ll b$, with $\mathsf F(q\mid b)=+\infty$ otherwise. If every nontrivial change is strictly costly, then $\lambda_{\mathrm F}>0$. Conversely, \cref{eq:axiomatic-filtering-cost} satisfies the axioms.
\end{proposition}

\begin{proof}
Define $g(t)=\ell_{\mathrm F}(e^t)$. Sequential consistency implies
\begin{align*}
g(s+t)
=
g(s)+g(t)
\end{align*}
Continuity therefore gives $g(t)=\lambda_{\mathrm F}t$ for some constant $\lambda_{\mathrm F}$, so
\begin{align*}
\ell_{\mathrm F}(r)
=
\lambda_{\mathrm F}\log r
\end{align*}
Substitution into \cref{eq:filter-locality} yields \cref{eq:axiomatic-filtering-cost}. Aggregate nonnegativity rules out $\lambda_{\mathrm F}<0$, and strict cost for every nontrivial change implies $\lambda_{\mathrm F}>0$. The converse follows from the additivity of the logarithm and the nonnegativity of relative entropy.
\end{proof}

The local quantity $\ell_{\mathrm F}(q(x)/b(x))$ is an accounting increment rather than a stand-alone expenditure. It is negative for types that are depleted relative to the baseline and positive for types that are enriched. Material balance links these terms, and the aggregate cost is nonnegative. Applied to the first stage, Proposition~\ref{prop:axiomatic-filtering} gives
\begin{align}\label{eq:first-stage-axiomatic}
\mathsf F(P_\xi\mid\phi)
=
\lambda_{\mathrm F}\KL(P_\xi\Vert\phi)
\end{align}
It also explains the support restriction $P_\xi\ll\phi$: a filter cannot deliver a characteristic that never appears in its ambient input.

\subsubsection{Axioms for sorting}

After filtering, the incoming stock has composition $p=P_\xi$. Before any deliberate sorting, random filling would place composition $p$ in every box, producing the independent joint distribution $p\otimes\mu$. A target joint distribution $P$ instead places composition $q_\theta$ in box $\theta$, subject to \cref{eq:sorting-balance}. Write $\mathsf S(P\mid p,\mu)$ for the cost per unit mass of implementing this allocation.

The sorting technology obeys parallel axioms.

\begin{enumerate}
\item \textit{Capacity and material conservation.} Box sizes remain $\mu$, the total stock remains $p$, and sorting changes only the assignment of particles to boxes. Thus the admissible targets are precisely the joint distributions with marginals $(p,\mu)$

\item \textit{Box separability and replication.} The cost of a collection of boxes is the capacity-weighted average of their within-box costs. Splitting one box into two smaller boxes with the same composition does not change total cost. Hence, for a type-independent local schedule $\ell_{\mathrm S}$,
\begin{align}\label{eq:sorting-locality}
\mathsf S(P\mid p,\mu)
=
\sum_{\theta\in\Theta}\mu(\theta)
\sum_{x:q_\theta(x)>0}q_\theta(x)
\ell_{\mathrm S}\left(\frac{q_\theta(x)}{p(x)}\right)
\end{align}
The ratio $q_\theta(x)/p(x)$ measures how strongly type $x$ is overrepresented or underrepresented in box $\theta$ relative to random filling

\item \textit{Hierarchical consistency.} A coarse sort followed by a fine sort costs the same as the corresponding direct sort. At the particle level this again requires
\begin{align}\label{eq:sorting-composition}
\ell_{\mathrm S}(rs)
=
\ell_{\mathrm S}(r)
+
\ell_{\mathrm S}(s)
\end{align}
For example, if fine boxes are indexed by $(g,h)$, then
\begin{align*}
\frac{P(\xi=x\mid g,h)}{p(x)}
=
\frac{P(\xi=x\mid g)}{p(x)}
\frac{P(\xi=x\mid g,h)}{P(\xi=x\mid g)}
\end{align*}
so the direct enrichment factor is the product of the coarse and within-group enrichment factors

\item \textit{Regularity and null sorting.} The local schedule is continuous, random filling has zero cost, and every non-independent allocation has nonnegative aggregate cost
\end{enumerate}

\begin{proposition}[Axiomatic sorting cost]\label{prop:axiomatic-sorting}
The sorting axioms imply that there is a constant $\lambda_{\mathrm S}\geq0$ such that
\begin{align}\label{eq:axiomatic-sorting-cost}
\mathsf S(P\mid p,\mu)
&=
\lambda_{\mathrm S}
\sum_{\theta\in\Theta}\mu(\theta)
\KL(q_\theta\Vert p)
\\
&=
\lambda_{\mathrm S}
\KL(P_{\xi,\theta}\Vert p\otimes\mu)
\\
&=
\lambda_{\mathrm S}\I_P(\xi;\theta)
\end{align}
If every non-independent allocation is strictly costly, then $\lambda_{\mathrm S}>0$. Conversely, \cref{eq:axiomatic-sorting-cost} satisfies the sorting axioms.
\end{proposition}

\begin{proof}
The same continuous multiplicative Cauchy equation used in Proposition~\ref{prop:axiomatic-filtering} gives
\begin{align*}
\ell_{\mathrm S}(r)
=
\lambda_{\mathrm S}\log r
\end{align*}
Substituting into \cref{eq:sorting-locality} gives
\begin{align*}
\mathsf S(P\mid p,\mu)
&=
\lambda_{\mathrm S}
\sum_{\theta,x}
P(x,\theta)
\log\frac{P(x\mid\theta)}{p(x)}
\\
&=
\lambda_{\mathrm S}
\sum_{\theta,x}
P(x,\theta)
\log\frac{P(x,\theta)}{p(x)\mu(\theta)}
\end{align*}
which is relative entropy from the random-fill allocation $p\otimes\mu$ and therefore mutual information.
\end{proof}

The same quantity has the symmetric representation
\begin{align}\label{eq:sorting-two-directions}
\I_P(\xi;\theta)
&=
\sum_{\theta\in\Theta}\mu(\theta)
\KL(P_{\xi\mid\theta}\Vert P_\xi)
\\
&=
\sum_{x\in X}P_\xi(x)
\KL(P_{\theta\mid\xi=x}\Vert\mu)
\end{align}
The first line compares each box's composition with the incoming stock. The second compares the destination frequencies of each particle type with the available box capacities. Thus mutual information is exactly the excess sorting needed to create dependence while preserving both marginals.

\subsubsection{The two-stage cost}

The final substantive assumption is \textit{stage separability}: filtering resources and sorting resources add, and any interaction between the stages is summarized by the intermediate composition $P_\xi$. Combining Propositions~\ref{prop:axiomatic-filtering} and~\ref{prop:axiomatic-sorting} gives
\begin{align}\label{eq:general-two-stage-cost}
C(P;\phi,\mu)
=
\lambda_{\mathrm F}\KL(P_\xi\Vert\phi)
+
\lambda_{\mathrm S}\I_P(\xi;\theta)
\end{align}
The paper's parameterization is
\begin{align}\label{eq:price-parameterization}
\lambda_{\mathrm S}
=
\lambda
\qquad
\lambda_{\mathrm F}
=
\eta\lambda
\end{align}
which reproduces \cref{eq:two-ways-cost}. In this microfoundation, $\lambda$ is the unit price of sorting and $\eta$ is the unit price of filtering relative to sorting.

The axioms characterize the two functional forms only up to the two nonnegative scale parameters $(\lambda_{\mathrm F},\lambda_{\mathrm S})$. They do not imply $\eta\leq1$. The maintained restriction $\eta\in[0,1]$ is therefore an additional technological restriction---equivalently, filtering is assumed to be weakly cheaper per unit than sorting---and is also the restriction inherited from the primitive Monty Hall accounting. A stand-alone filtering-and-sorting model could instead take $(\lambda_{\mathrm F},\lambda_{\mathrm S})$ as its primitive parameters.

When the two unit prices are equal, $\lambda_{\mathrm F}=\lambda_{\mathrm S}=\lambda$, the relative-entropy chain rule gives
\begin{align}\label{eq:equal-price-chain-rule}
\lambda\KL(P_\xi\Vert\phi)
+
\lambda\I_P(\xi;\theta)
=
\lambda\KL(P_{\xi,\theta}\Vert\phi\otimes\mu)
=
\lambda\E_{\theta\sim\mu}
\left[
\KL(P_{\xi\mid\theta}\Vert\phi)
\right]
\end{align}
Thus the endpoint $\eta=1$ means that filtering and sorting have the same unit price, not that state dependence is costless. At $\eta=0$, ambient filtering is free and only sorting the chosen marginal across states is costly.

\begin{remark}[Scope of the characterization]\label{rem:axiomatic-scope}
The substantive restrictions are relative-frequency locality, accounting per unit of delivered output, box separability, and exact sequential consistency. Relabeling, continuity, and nonnegativity by themselves do not select relative entropy from the many possible divergence measures. The filtering construction also abstracts from gross throughput: it prices the composition of one unit of delivered particles but not the number of ambient particles discarded along the way. A technology with throughput costs, fixed setup costs, congestion, or interactions across boxes would generally add further terms. The propositions characterize the KL components within the ideal composition-only class stated by the axioms.
\end{remark}

\subsection{Interpreting the ambient distribution}\label{sec:phi-interpretation}

In the literal Monty Hall model, $\phi(x)=n_x/J$. More generally, $\phi$ is the reference distribution before payoff-directed filtering. It can combine physical availability, visibility, shelf position, screen position, salience, or the frequency with which an option enters a consumer's initial consideration. Thus $\phi$ need not be uniform and need not equal literal market share. It is better interpreted as \textit{ambient exposure measured in units relevant for information acquisition}.

This interpretation imposes discipline. The reference must be determined before the agent's information policy, must place mass only on feasible characteristics, and must describe what the filtering technology treats as readily available. It should not absorb utility differences that the model is meant to estimate. In a randomized interface experiment, $\phi$ may be induced by the randomization. In an observational market, it may be a parameterized exposure index. The literal $1/J$ case is therefore a useful benchmark, not a defining restriction.

\subsection{A two-contract health-insurance example}\label{sec:insurance-example}

Consider two contract characteristics,
\begin{align*}
X=\{H,L\}
\end{align*}
where $H$ denotes high coverage and $L$ denotes low coverage, and two health states,
\begin{align*}
\Theta=\{h,\ell\}
\end{align*}
where $h$ denotes high risk and $\ell$ denotes low risk. Suppose the two states are equally likely and utility is one when coverage matches risk and zero otherwise
\begin{align}\label{eq:insurance-match-utility}
\mu(h)
&=
\mu(\ell)
=
\frac12
\\
u(H,h)
&=
u(L,\ell)=1
\\
u(H,\ell)
&=
u(L,h)=0
\end{align}
There are two arbitrary contract labels, one hiding $H$ and the other hiding $L$. Thus
\begin{align*}
J=2
\qquad
n_H=n_L=1
\qquad
\phi(H)=\phi(L)=\frac12
\end{align*}
and the Monty Hall prior is uniform over the two assignments $(H,L)$ and $(L,H)$. The economic menu is fixed: it always contains one high-coverage contract and one low-coverage contract. Only their assignment to labels is initially unknown.

By symmetry, an optimal reduced policy has the form
\begin{align}\label{eq:insurance-symmetric-policy}
P(\xi=H\mid h)
&=
P(\xi=L\mid\ell)
=
q
\\
P(\xi=L\mid h)
&=
P(\xi=H\mid\ell)
=
1-q
\end{align}
for some $q\in[0,1]$. Every such policy has
\begin{align*}
P_\xi(H)=P_\xi(L)=\frac12
\qquad
P_\xi=\phi
\end{align*}
so marginal filtering costs are zero. The example therefore isolates sorting. Expected utility is $q$, while mutual information is
\begin{align}\label{eq:insurance-mutual-information}
\I_P(\xi;\theta)
=
q\log(2q)
+
(1-q)\log\bigl(2(1-q)\bigr)
\end{align}
The reduced problem becomes
\begin{align}\label{eq:insurance-one-dimensional-problem}
\max_{q\in[0,1]}
\left\{
q
-
\lambda
\left[
q\log(2q)
+
(1-q)\log\bigl(2(1-q)\bigr)
\right]
\right\}
\end{align}
and its unique solution is
\begin{align}\label{eq:insurance-match-probability}
q^*
=
\frac{e^{1/\lambda}}{1+e^{1/\lambda}}
\end{align}
The parameter $\eta$ drops out because the desired marginal already equals the ambient composition. With no sorting, $q=1/2$, selected coverage is independent of risk, expected utility is $1/2$, and information cost is zero. Perfect sorting, $q=1$, raises expected utility to one and costs $\lambda\log2$. As $\lambda$ falls, the agent more closely matches high coverage to high risk and low coverage to low risk.

The example clarifies what the model treats as costly. Learning that high coverage is common would not help: high and low coverage are equally common, and changing their aggregate frequencies cannot raise expected utility. The gain comes entirely from making the selected characteristic covary with the health state. This is precisely the operation measured by $\I_P(\xi;\theta)$.

\subsubsection{Why a one-message cost does not separately price the health state}

The comparison with the standard Mat\v{e}jka--McKay formulation is especially sharp in this example. Let $M\in\{0,1\}$ encode the assignment of coverage characteristics to labels, let $a\in\{1,2\}$ be the recommended label, and suppose $M$ and $\theta$ are independent fair binary variables. Under perfect matching, the recommended label is the parity of the assignment and health state. It follows that
\begin{align}\label{eq:insurance-xor-information}
\I(a;M)
&=
0
\\
\I(a;\theta)
&=
0
\\
\I(a;(M,\theta))
&=
\log2
\end{align}
The recommendation reveals neither input by itself, even though it is completely determined by the pair.

Now compare two versions of a one-message Shannon problem. If the health state is already known to the consumer, implementing perfect matching requires
\begin{align}\label{eq:mm-known-state-cost}
\lambda\I(a;M\mid\theta)
=
\lambda\log2
\end{align}
If the health state is initially unknown and is included in the payoff state, the same rule requires
\begin{align}\label{eq:mm-unknown-state-cost}
\lambda\I(a;(M,\theta))
=
\lambda\log2
\end{align}
Thus adding uncertainty about $\theta$ does not raise the one-message cost. This does not mean that the Shannon cost is literally zero. It means that the model charges for the information contained in the final recommendation and does not separately charge for the number or nature of the latent inputs that must be combined to produce it. In this precise sense, learning $\theta$ is free relative to the counterfactual in which $\theta$ was already known.

The filtering-and-sorting decomposition distinguishes the two cases. If $\theta$ is observed without cost, then conditional filtering selects $H$ in state $h$ and $L$ in state $\ell$, at average cost
\begin{align}\label{eq:insurance-known-state-two-stage-cost}
\eta\lambda
\E_{\theta\sim\mu}
\left[
\KL(P_{\xi\mid\theta}\Vert\phi)
\right]
=
\eta\lambda\log2
\end{align}
If $\theta$ must also be learned, perfect matching costs
\begin{align}\label{eq:insurance-unknown-state-two-stage-cost}
\eta\lambda\log2
+
(1-\eta)\lambda\log2
=
\lambda\log2
\end{align}
The extra cost of initially unknown health risk is therefore $(1-\eta)\lambda\log2$. The two models agree when filtering and sorting have the same unit price, $\eta=1$, but otherwise the one-message model cannot represent the distinction.

\subsubsection{Comparison with Brown and Jeon}

\textcite{brown2024endogenous} apply the Mat\v{e}jka--McKay framework to insurance choice by dividing utility into components that are initially known and plan-specific components that are initially unknown. Premiums and some plan characteristics are known. Expected out-of-pocket costs, plan risk, and other difficult-to-observe characteristics enter an initially unknown utility vector. Their maintained prior makes the unknown components independent across plans. Their empirical specification also includes preference heterogeneity that is known to the consumer, including an idiosyncratic taste shock and observed interactions between diagnoses and insurers. Consumers can therefore act on those dimensions of heterogeneity without first paying to learn which type they are.

The distinction is not that Brown and Jeon ignore health uncertainty. Expected out-of-pocket costs and plan risk require consumers to reason about contract terms and potential health shocks. Rather, these ingredients are compressed into plan-specific utility draws before the information problem is solved. A common health state, the characteristics of the contracts, and the matching of one to the other do not receive separate information prices. As \cref{eq:mm-known-state-cost,eq:mm-unknown-state-cost} show, a direct-recommendation cost can be unchanged when uncertainty about a common state is added to the problem. If a component of health risk changes only the common level of utility and not the ranking of labels, it is entirely irrelevant to the action. If it changes rankings jointly with contract characteristics, it can affect the recommendation without generating a separate state-learning charge.

This matters for Brown and Jeon's restricted-menu counterfactual. They find that removing low-mean-utility plans initially raises welfare by reducing research costs and mistakes, but that removing too many plans eventually lowers welfare because consumers with heterogeneous preferences lose plans that fit them well. Part of the heterogeneity producing these good matches is treated as known to the consumer. In that case, the consumer can exploit a finely differentiated menu without paying a separate cost to determine which health or preference state makes a specialized plan attractive.

In the present model, additional specialized contracts create option value only to the extent that the consumer can sort them toward the states in which they are useful. That sorting costs $\lambda\I_P(\xi;\theta)$. When state sorting is expensive, fine differentiation is less valuable than it would be for an otherwise identical consumer who knows her type for free. Consequently, the welfare-maximizing menu can be smaller, and the point at which further plan removal becomes harmful can occur after more plans have been removed. This is a qualitative sensitivity argument rather than a signed theorem for every menu change: deleting a plan may also change ambient exposure $\phi$, remove a characteristic from its support, or alter utility directly.

\subsubsection{Independent draws versus fixed menu composition}

The joint prior over contract characteristics creates a second difference. To see it, replace the Monty Hall prior with independent draws
\begin{align}\label{eq:insurance-iid-prior}
\xi_1,\xi_2
\mathrel{\overset{\mathrm{iid}}{\sim}}
\phi
\qquad
\phi(H)=\phi(L)=\frac12
\end{align}
The four menus
\begin{align*}
(H,H)
\qquad
(H,L)
\qquad
(L,H)
\qquad
(L,L)
\end{align*}
then each occur with probability $1/4$. Half of the prior probability is placed on duplicate menus. In the events $(H,H)$ and $(L,L)$,
\begin{align}\label{eq:duplicate-menu-indifference}
u(\xi_1,\theta)
=
u(\xi_2,\theta)
\qquad
\text{for both }\theta\in\{h,\ell\}
\end{align}
so neither the assignment of characteristics nor the health state affects which label is better. They may still affect the common level of realized utility, but they are irrelevant to the choice. A direct-recommendation policy can choose a fixed label in these events and acquire no action-relevant information.

This lowers the value assigned to information. Under the fixed Monty Hall menu, full information always produces a matching contract, so its expected payoff is one. Under independent draws,
\begin{align}\label{eq:iid-full-information-payoff}
\E\left[
\max_{j\in\{1,2\}}u(\xi_j,\theta)
\right]
=
1-\frac14
=
\frac34
\end{align}
because, conditional on either health state, both contracts fail to match with probability $1/4$. An uninformed choice earns $1/2$ under either prior. Hence perfect information raises expected utility by $1/2$ under fixed composition but only by $1/4$ under independent draws. The independent prior makes information about states and characteristics less valuable because it assigns positive probability to menus on which no information can improve the action.

With $J$ independently drawn binary contracts, the full-information probability of finding a match is
\begin{align}\label{eq:iid-menu-size-match}
1-2^{-J}
\end{align}
and the probability that the menu contains both coverage types is $1-2^{1-J}$. Menu size therefore mechanically creates expected characteristic diversity. Under a continuous independent prior, exact duplicates have probability zero, so almost every additional plan is potentially a new idiosyncratic match. Under fixed-composition uncertainty, by contrast, additional labels may merely replicate characteristics already present. Removing labels while retaining at least one contract of each relevant type need not destroy any full-information matching opportunity; removing the last contract of a type does.

This distinction can materially affect counterfactual interpretation. An independent-draw prior may understate the baseline value of learning a common health state because some prior menus contain no useful tradeoff. At the same time, it can make the welfare loss from aggressive menu reduction appear larger by treating each additional plan as another independent opportunity to obtain a characteristic favored by some consumer. A downturn in welfare as plans are removed may then combine two conceptually different forces: genuine loss of preference-relevant variety and a mechanically induced reduction in the probability that the menu spans the relevant characteristics. The fixed-composition model separates three interventions that a plan count alone confounds: deleting redundant labels, changing ambient exposure $\phi$, and removing a characteristic from the feasible support.

The binary calculation is an illustration of the joint-prior restriction, not a literal description of Brown and Jeon's empirical simulation. Their counterfactual removes observed plans with low predicted mean utility and recomputes the stakes and information costs; it does not redraw insurance contracts. The calculation instead shows why conclusions about how far a menu can be simplified may be sensitive to whether uncertainty concerns a fixed but hidden composition or independent plan-specific utility draws. Determining the numerical effect on their counterfactuals would require re-estimating the model under a correlated or fixed-composition prior together with a separate cost for learning health states.

\section{Solution and Implications}\label{sec:results}

Assume throughout this section that $u$ is finite, $\mu(\theta)>0$ for every $\theta$, $\phi(x)>0$ for every $x$, $\lambda>0$, and $\eta\in(0,1]$.

\subsection{Existence, uniqueness, and support preservation}

\begin{theorem}[Existence, uniqueness, and full support]\label{th:existence-uniqueness}
Problem \cref{eq:reduced-problem} has a unique solution $P^*$. Moreover,
\begin{align}\label{eq:full-support}
p^*(x\mid\theta)>0
\qquad
\text{for every }x\in X\text{ and }\theta\in\Theta
\end{align}
where $p^*(x\mid\theta)=P^*(\xi=x\mid\theta)$. Consequently $P^*_\xi$ and $\phi$ are mutually absolutely continuous.
\end{theorem}

\begin{proof}
The feasible set is a compact convex product of finite-dimensional probability simplexes. With the convention $0\log 0=0$, the objective is continuous, so a maximizer exists.

For a feasible $P$, write $p_\theta(x)=P(\xi=x\mid\theta)$. For fixed $P_\theta=\mu$, mutual information is convex in the collection $(p_\theta)_{\theta\in\Theta}$. The first expression in \cref{eq:two-ways-cost} contains
\begin{align*}
\eta\lambda\sum_{\theta\in\Theta}\mu(\theta)
\KL(p_\theta\Vert\phi)
\end{align*}
which is strictly convex in the collection of conditional probabilities because $\eta\lambda>0$ and relative entropy is strictly convex. The total cost is therefore strictly convex, and expected utility is linear. The objective is strictly concave and its maximizer is unique.

Suppose $p_\theta(x)=0$ at a maximizer. Shift probability $\varepsilon$ to $x$ from any positive-probability characteristic in that state. The utility change is of order $\varepsilon$, whereas the right derivative of the negative entropy term at zero is positive infinity because the derivative of $-p\log p$ diverges as $p\downarrow0$. For sufficiently small $\varepsilon$, the perturbation raises the objective, a contradiction. Thus \cref{eq:full-support} holds. Summing over states gives $P_\xi(x)>0$ whenever $\phi(x)>0$.
\end{proof}

The conclusion is stronger than the feasibility restriction $P_\xi\ll\phi$. Feasibility says the agent cannot create characteristics absent from the ambient environment. The theorem says the agent also never completely removes a characteristic that is present. In finite spaces, $P^*_\xi\sim\phi$.

At $\eta=0$, existence remains but uniqueness and full support can fail. This is the familiar sparse-consideration case in Shannon rational inattention \parencite{caplin2019rational}. Any positive $\eta$ acts as an entropy barrier against eliminating an ambient characteristic.

\subsection{The choice rule}

\begin{proposition}[Weighted logit]\label{prop:weighted-logit}
The unique optimum satisfies
\begin{align}\label{eq:weighted-logit}
p^*(x\mid\theta)
=
\frac{
\phi(x)^\eta p^*(x)^{1-\eta}e^{u(x,\theta)/\lambda}
}{
\sum_{y\in X}\phi(y)^\eta p^*(y)^{1-\eta}e^{u(y,\theta)/\lambda}
}
\end{align}
where $p^*(x)=P^*_\xi(x)$. Conversely, any feasible full-support distribution whose conditional and marginal probabilities satisfy \cref{eq:weighted-logit} is the unique optimum.
\end{proposition}

\begin{proof}
Write $p_\theta(x)=P(\xi=x\mid\theta)$ and $p(x)=P_\xi(x)=\sum_\theta\mu(\theta)p_\theta(x)$. The first-order condition for $p_\theta(x)$, after absorbing terms common across $x$ into the multiplier for $\sum_xp_\theta(x)=1$, is
\begin{align*}
u(x,\theta)
-
\eta\lambda\log\frac{p(x)}{\phi(x)}
-
\lambda\log\frac{p_\theta(x)}{p(x)}
=
\chi(\theta)
\end{align*}
Rearranging and normalizing over $x$ gives \cref{eq:weighted-logit}. Strict concavity makes the first-order condition sufficient and makes its feasible solution unique.
\end{proof}

The rule balances three forces. Utility contributes $u/\lambda$. Ambient exposure contributes $\eta\log\phi$. The agent's unconditional behavior contributes $(1-\eta)\log p$. In log-odds form,
\begin{align}\label{eq:log-odds}
\log\frac{p(x\mid\theta)}{p(y\mid\theta)}
=
\frac{u(x,\theta)-u(y,\theta)}{\lambda}
+
\eta\log\frac{\phi(x)}{\phi(y)}
+
(1-\eta)\log\frac{p(x)}{p(y)}
\end{align}
This equation is the main bridge between the model and data.

The endpoints have transparent meanings under the filtering-and-sorting interpretation. At $\eta=0$, ambient filtering is free and only sorting across states is costly, so the rule uses the endogenous marginal $p$ as in the standard state--action model. At $\eta=1$, filtering and sorting have the same unit price; the chain rule in \cref{eq:equal-price-chain-rule} then makes the rule use the exogenous reference $\phi$. For intermediate $\eta$, filtering is cheaper than sorting and the geometric mean $\phi^\eta p^{1-\eta}$ interpolates between the two references.

Because the objective is strictly concave, computation requires no search over multiple equilibria. One may maximize \cref{eq:reduced-problem} directly on the product of simplexes or solve \cref{eq:weighted-logit} together with
\begin{align*}
p(x)
=
\sum_{\theta\in\Theta}\mu(\theta)p(x\mid\theta)
\end{align*}

\subsection{Strictly dominated characteristics}

The model can assign positive probability to a strictly dominated characteristic without treating it as a deliberately chosen action. The agent chooses a recommendation label under uncertainty; $\xi$ records what lies behind the chosen label.

\begin{corollary}[Dominated characteristics]\label{cor:dominated}
Suppose $x^+$ strictly dominates $x^-$
\begin{align*}
u(x^+,\theta)>u(x^-,\theta)
\qquad
\text{for every }\theta\in\Theta
\end{align*}
If $\phi(x^-)>0$, then
\begin{align*}
p^*(x^-\mid\theta)>0
\qquad
p^*(x^-)>0
\end{align*}
If the utility gap is constant, $u(x^+,\theta)-u(x^-,\theta)=c>0$, then
\begin{align}\label{eq:dominated-odds}
\frac{p^*(x^+)}{p^*(x^-)}
=
\frac{\phi(x^+)}{\phi(x^-)}
e^{c/(\eta\lambda)}
\end{align}
\end{corollary}

\begin{proof}
Positivity follows from \cref{th:existence-uniqueness}. With a constant gap, \cref{eq:log-odds} implies that the conditional odds are independent of $\theta$ and hence equal the marginal odds
\begin{align*}
\frac{p^*(x^+)}{p^*(x^-)}
=
\left(\frac{\phi(x^+)}{\phi(x^-)}\right)^\eta
\left(\frac{p^*(x^+)}{p^*(x^-)}\right)^{1-\eta}
e^{c/\lambda}
\end{align*}
which yields \cref{eq:dominated-odds}.
\end{proof}

The dominated characteristic is rare when the utility loss is large or information is cheap, but it is not assigned probability zero. Completely eliminating it would require an arbitrarily sharp filter. Equation \cref{eq:dominated-odds} also supplies an identifying restriction for the product $\eta\lambda$ when a constant utility gap and relative exposure are known.

\subsection{A utility-free rule for a designer}

The ambient distribution is a natural policy lever when a designer controls pre-choice exposure without directly changing payoffs. A digital platform can alter candidate-set sampling, search placement, or display frequency. A retailer can reallocate shelf space or inventory, and a benefits portal or comparison site can change how frequently plans appear. In the literal menu model, these interventions change the number of labels carrying each characteristic. Under the broader interpretation of $\phi$, they change attention-weighted availability. Because total exposure is fixed, increasing some components requires decreasing others.

A full structural approach would estimate utility, information costs, preference heterogeneity, and ambient exposure, then solve the designer's problem. That approach is useful for large counterfactuals or interventions with additional constraints, but it is demanding and can be weakly identified. The rule below instead uses only the current exposure distribution and the marginal distribution of observed choices. It gives a transparent welfare-improving direction that can be implemented gradually or tested experimentally.

Let
\begin{align*}
V(\phi)
=
\max_{P\in\calQ}
\left\{
\E_P[u]
-
\eta\lambda\KL(P_\xi\Vert\phi)
-
\lambda\I_P(\xi;\theta)
\right\}
\end{align*}
Let $p^\phi$ denote the marginal distribution of the chosen characteristic under the unique optimum.

\begin{proposition}[Shift exposure toward revealed demand]\label{prop:designer}
For any direction $h\in\R^{|X|}$ with $\sum_xh(x)=0$, the directional derivative is
\begin{align}\label{eq:designer-derivative}
DV(\phi)[h]
=
\eta\lambda\sum_{x\in X}\frac{p^\phi(x)}{\phi(x)}h(x)
\end{align}
If $h(x)$ has the same sign as $p^\phi(x)-\phi(x)$ for every $x$, with $h(x)[p^\phi(x)-\phi(x)]>0$ for at least one $x$, then $DV(\phi)[h]>0$. In particular, the update
\begin{align*}
\phi_\varepsilon
=
(1-\varepsilon)\phi+\varepsilon p^\phi
\end{align*}
raises welfare for every $\varepsilon\in(0,1]$ unless $p^\phi=\phi$.
\end{proposition}

\begin{proof}
The envelope theorem leaves only the direct effect of $\phi$ on the KL term
\begin{align*}
DV(\phi)[h]
=
\eta\lambda\sum_xp^\phi(x)\frac{h(x)}{\phi(x)}
\end{align*}
Because $\sum_xh(x)=0$, this equals
\begin{align*}
\eta\lambda\sum_x
\left(
\frac{p^\phi(x)}{\phi(x)}-1
\right)h(x)
\end{align*}
which is positive under the sign condition. For $h=p^\phi-\phi$, the derivative is
\begin{align*}
\eta\lambda
\left[
\sum_x\frac{p^\phi(x)^2}{\phi(x)}-1
\right]
\geq
0
\end{align*}
by Cauchy--Schwarz, with equality only when $p^\phi=\phi$.

The stronger claim for $\phi_\varepsilon$ does not rely on a local approximation. Hold the information policy at its old optimum. Convexity of relative entropy in its second argument gives
\begin{align*}
\KL(p^\phi\Vert\phi_\varepsilon)
\leq
(1-\varepsilon)\KL(p^\phi\Vert\phi)
\end{align*}
The old policy therefore delivers the same expected utility and state information while requiring strictly less information about characteristics whenever $p^\phi\neq\phi$. Reoptimization can only raise welfare further.
\end{proof}

The coordinatewise prescription is simple: subject to the exposure budget, increase $\phi(x)$ if and only if $p^\phi(x)>\phi(x)$. A characteristic with $p^\phi(x)>\phi(x)$ is selected more often than ambient exposure alone would predict, so the agent is spending information to find it. Making it more common reduces that cost. A characteristic with $p^\phi(x)<\phi(x)$ is filtered out, so reducing its exposure also saves information. Moving $\phi$ toward $p^\phi$ makes the agent's revealed policy cheaper to implement.

The result is utility-free, not because utility is irrelevant, but because it has already done its work. The optimal policy embodies the agent's tradeoff between payoff gains and information acquisition. At that optimum, the envelope theorem removes the first-order effect of the agent's behavioral response; only the direct reduction in filtering cost remains. This is why the designer need not reconstruct the utility function to obtain a welfare-improving direction. Structural estimation remains necessary for global counterfactuals, distributional effects, or constraints not represented by $\phi$, but it is unnecessary for this robust exposure adjustment.

For a constant-gap dominated pair, \cref{eq:dominated-odds} implies
\begin{align*}
\frac{p^\phi(x^-)}{\phi(x^-)}
<
\frac{p^\phi(x^+)}{\phi(x^+)}
\end{align*}
Thus a marginal shift of exposure from the dominated characteristic to its dominator raises welfare. In a two-characteristic comparison this also implies $p^\phi(x^-)<\phi(x^-)$, so the componentwise rule reduces the dominated characteristic. With additional alternatives, domination alone ranks the two likelihood ratios but need not force the dominated characteristic's ratio below one; the correct general rule remains to shift exposure from lower $p^\phi/\phi$ toward higher $p^\phi/\phi$. At every positive $\phi(x^-)$ the agent still selects $x^-$ with positive probability, but a designer can move it to the boundary by changing the environment rather than by assuming perfect attention.

\section{Identification and Estimation}\label{sec:empirics}

\subsection{The collinearity problem}

The log-odds equation contains the composite terms
\begin{align*}
\eta\log\frac{\phi(x)}{\phi(y)}
\qquad
(1-\eta)\log\frac{p(x)}{p(y)}
\end{align*}
Within a single market, an unrestricted $\phi$ behaves much like a collection of alternative-specific intercepts whose coefficients depend on $\eta$. A change in $\eta$ can be offset by a change in the ratios of $\phi$, especially when utility already contains flexible characteristic fixed effects. At $\eta=0$, $\phi$ disappears and is not identified at all. When $\eta$ is close to zero, it is weakly identified.

This is why jointly estimating an unrestricted $(\eta,\phi)$ from one cross section is unattractive even when the nonlinear model is formally identified. The objective may have a long, nearly flat ridge. Estimating $\eta$ first from entry variation removes the main rotation between the exponent on $\phi$ and the exponent on the marginal choice probability $p$.

\subsection{Identification through entry}\label{sec:entry}

Consider two environments $t\in\{0,1\}$. Characteristic $x_j$ enters at $t=1$, while incumbents $x_1$ and $x_2$ remain available. Let $X_t$ be the set of available characteristics, let $p_t(x\mid\theta)$ and $p_t(x)$ be the conditional and marginal choice probabilities, and let $\phi_t$ and $u_t$ denote ambient exposure and utility. Assume
\begin{enumerate}
\item incumbent utilities are stable, so $u_1(x_k,\theta)=u_0(x_k,\theta)$ for $k=1,2$
\item the information parameters $(\eta,\lambda)$ and state distribution are stable
\item entry does not change the incumbents' relative ambient exposure
\begin{align}\label{eq:stable-phi-ratio}
\frac{\phi_1(x_1)}{\phi_1(x_2)}
=
\frac{\phi_0(x_1)}{\phi_0(x_2)}
\end{align}
\end{enumerate}
A sufficient condition for the last assumption is a stable exposure index $v(x)>0$ with
\begin{align*}
\phi_t(x)
=
\frac{v(x)}{\sum_{y\in X_t}v(y)}
\end{align*}
Entry then changes the normalization of $\phi$ but not the ratio between any two incumbents. The assumption can be imposed only for selected pairs for which the entrant plausibly does not alter relative placement or salience.

Define the entry changes in conditional and marginal log odds
\begin{align}
\Delta_\theta^{\mathrm{cond}}(x_1,x_2)
&=
\log\frac{p_1(x_1\mid\theta)}{p_1(x_2\mid\theta)}
-
\log\frac{p_0(x_1\mid\theta)}{p_0(x_2\mid\theta)}
\label{eq:conditional-entry-change}
\\
\Delta^{\mathrm{marg}}(x_1,x_2)
&=
\log\frac{p_1(x_1)}{p_1(x_2)}
-
\log\frac{p_0(x_1)}{p_0(x_2)}
\label{eq:marginal-entry-change}
\end{align}

\begin{proposition}[Entry identification]\label{prop:entry-identification}
Under the three entry assumptions,
\begin{align}\label{eq:entry-moment}
\Delta_\theta^{\mathrm{cond}}(x_1,x_2)
=
(1-\eta)\Delta^{\mathrm{marg}}(x_1,x_2)
\qquad
\text{for every }\theta\in\Theta
\end{align}
If $\Delta^{\mathrm{marg}}(x_1,x_2)\neq0$, then
\begin{align}\label{eq:eta-entry}
\eta
=
1-
\frac{\Delta_\theta^{\mathrm{cond}}(x_1,x_2)}
{\Delta^{\mathrm{marg}}(x_1,x_2)}
\end{align}
Multiple states, entrants, and incumbent pairs over-identify $\eta$ and imply the testable restriction that $\Delta_\theta^{\mathrm{cond}}(x_1,x_2)$ is constant across $\theta$.
\end{proposition}

\begin{proof}
Apply \cref{eq:log-odds} in each environment and difference. Stable utility eliminates the utility gap, while \cref{eq:stable-phi-ratio} eliminates the change in relative ambient exposure. The remaining term is \cref{eq:entry-moment}. Equation \cref{eq:eta-entry} follows when the marginal-odds change is nonzero.
\end{proof}

The condition $\Delta^{\mathrm{marg}}\neq0$ matters. Entry that does not move the incumbents' relative marginal shares contains no information about $\eta$ through this moment. Likewise, a violation of \cref{eq:stable-phi-ratio} is observationally equivalent to a change in $\eta$ unless relative exposure is observed or otherwise restricted. The cross-state equality is therefore important as a specification check, not merely a source of precision.

One can estimate $\eta$ by GMM from moments
\begin{align}\label{eq:entry-gmm}
\E\left[
\Delta_\theta^{\mathrm{cond}}(x_1,x_2)
-(1-\eta)\Delta^{\mathrm{marg}}(x_1,x_2)
\right]
=
0
\end{align}
stacked across usable entries, pairs, and states. Sampling uncertainty in estimated shares should be carried into the second stage through stacked GMM, an influence-function correction, or a bootstrap that repeats both stages.

\subsection{Random coefficients and ambient exposure}

Let preferences vary with $\beta\sim F(\cdot\mid\gamma)$ and let utility be $u(x,\theta;\beta)$. For each type, the unique conditional probabilities solve
\begin{align}\label{eq:rc-type-ccp}
p_\beta(x\mid\theta)
=
\frac{
\phi(x)^\eta p_\beta(x)^{1-\eta}e^{u(x,\theta;\beta)/\lambda}
}{
\sum_{y\in X}
\phi(y)^\eta p_\beta(y)^{1-\eta}e^{u(y,\theta;\beta)/\lambda}
}
\end{align}
with
\begin{align*}
p_\beta(x)
=
\sum_{\theta\in\Theta}\mu(\theta)p_\beta(x\mid\theta)
\end{align*}
The observed conditional share is
\begin{align}\label{eq:integrated-share}
s(x\mid\theta)
=
\int p_\beta(x\mid\theta)
\,dF(\beta\mid\gamma)
\end{align}

The source of identification for $\phi$ is transparent at the type level. For any reference characteristic $y$ and $\eta>0$, \cref{eq:log-odds} implies
\begin{align}\label{eq:phi-inversion}
\log\frac{\phi(x)}{\phi(y)}
=
\frac1\eta
\left[
\log\frac{p_\beta(x\mid\theta)}{p_\beta(y\mid\theta)}
-
\frac{u(x,\theta;\beta)-u(y,\theta;\beta)}{\lambda}
-
(1-\eta)\log\frac{p_\beta(x)}{p_\beta(y)}
\right]
\end{align}
The right-hand side must be the same across states and preference types. Thus, once standard utility variation identifies $(\lambda,\gamma)$ and type-specific behavior is observed or recovered, \cref{eq:phi-inversion} identifies all relative values of $\phi$; the normalization $\sum_x\phi(x)=1$ identifies its level. Extra states and types supply over-identifying restrictions.

After estimating $\eta$ from entry, parameterize $\phi$ parsimoniously, for example through positive exposure indices $v(x;\alpha)$
\begin{align*}
\phi_t(x;\alpha)
=
\frac{v(x;\alpha)}{\sum_{y\in X_t}v(y;\alpha)}
\end{align*}
The normalization is essential because only relative exposure matters. For any trial $(\lambda,\gamma,\alpha)$, solve the type-specific strictly concave problems, integrate as in \cref{eq:integrated-share}, and compare predicted shares and micro moments with the data. Simulated maximum likelihood, simulated method of moments, or GMM are natural estimators; a mixed-logit share inversion can be used when the utility specification contains market-level mean utilities \parencite{train2009discrete}.

Under the usual discrete-choice requirements---utility shifters with sufficient independent variation, a scale normalization, a finite-dimensional identifiable $F(\cdot\mid\gamma)$, and excluded variation that moves exposure without moving utility or vice versa---the remaining parameters $(\lambda,\gamma,\phi)$ may be identified after $\eta$. The full vector $(\eta,\lambda,\gamma,\phi)$ may be over-identified because the entry moments, state-specific shares, substitution patterns, and micro moments all restrict the same parameters. This is a possibility under rank and exclusion conditions, not an automatic consequence of having many observed shares.

There is an important heterogeneity caveat. The entry identity \cref{eq:entry-moment} holds type by type. An odds ratio of integrated shares generally does not equal the integral of type-specific odds ratios. With latent random coefficients, the clean first-stage entry design therefore requires either individual panels, sufficiently informative observed types, or micro moments that recover type-specific substitution. Without such information, the entry restrictions should be imposed inside the joint simulated model rather than applied mechanically to aggregate shares.

\subsection{Detecting under-identification}

Several failures are especially relevant in this model.

\begin{enumerate}
\item \textit{Utility scale.} If $u(x,\theta;\beta)=z(x,\theta)'\beta$ and the scale of $\beta$ is unrestricted, only $\beta/\lambda$ is identified. One must fix a utility coefficient, express utility in known monetary units, normalize $\lambda$, or bring in information that fixes utility scale

\item \textit{Flexible exposure.} Unrestricted alternative-by-market values of $\phi$ can absorb utility intercepts and substitution patterns. A stable exposure index, randomized placement, measured impressions, or exclusion restrictions are needed to distinguish exposure from tastes

\item \textit{Excessive heterogeneity.} A high-dimensional $\gamma$ or nearly unrestricted mixing distribution can reproduce many share patterns and mimic variation in $\phi$. Panel choices and targeted micro moments are often more useful than additional aggregate shares. Regularization can stabilize estimation but does not create identification

\item \textit{Weak entry variation.} If $\Delta^{\mathrm{marg}}(x_1,x_2)$ is close to zero, \cref{eq:eta-entry} is weak even with a large sample. If $\eta$ is close to zero, $\phi$ is weakly identified in the second stage
\end{enumerate}

In implementation, estimate the Jacobian of the stacked moments with respect to $(\eta,\lambda,\gamma,\alpha)$ and inspect its smallest singular values and condition number. A nearly singular Jacobian, a nearly singular likelihood Hessian, estimates that move sharply across starting values, or flat profile likelihood and GMM objective functions are direct warnings. Profile each economically important parameter rather than relying only on local standard errors.

An over-identification $J$ test asks whether the model fits the extra moments; it does not by itself establish identification. A model can satisfy many redundant moments and still have a rank-deficient parameter map. When weak identification is plausible, report weak-identification-robust or set-valued confidence regions, reduce the dimension of $\gamma$ or $\phi$, and show which added exclusions or micro moments restore rank. These diagnostics should be performed before interpreting a precise optimizer as a precisely identified economic parameter.

\section{Conclusion}

The paper's point is that uncertainty about a state and uncertainty about what lies behind a choice label are economically different even when both are summarized by one recommendation. A finite Monty Hall model makes the distinction explicit and reduces exactly to a tractable problem over chosen characteristics and states.

The reduced information cost has two pieces: the work needed to filter an ambient distribution $\phi$ into the chosen marginal, and the work needed to match that marginal to states. This delivers a unique, full-support policy and the weighted logit rule
\begin{align*}
p(x\mid\theta)
\propto
\phi(x)^\eta p(x)^{1-\eta}e^{u(x,\theta)/\lambda}
\end{align*}
The same structure explains dominated realized characteristics, yields a designer rule based only on the gap between exposure and choice, and generates entry moments that identify the relative information-cost parameter before estimating the harder exposure and preference components.

The abstract interpretation of $\phi$ is important. It is not necessarily uniform random choice. It is the ambient distribution from which payoff-directed attention filters. Treating it this way makes the model useful beyond literal menus while also clarifying what empirical variation is required to distinguish ambient exposure from tastes. The health-insurance example further shows why menu simplification must distinguish redundant labels, changes in characteristic exposure, and the elimination of characteristics needed in particular states; those interventions can look identical when summarized only by the number of plans.

\printbibliography

\end{document}